\documentclass[10pt, conference, compsocconf]{IEEEtran}

\usepackage{graphicx}
\usepackage{amsmath}

\ifCLASSOPTIONcompsoc
\usepackage[caption=false,font=normalsize,labelfont=sf,textfont=sf]{subfig}
\else \usepackage[caption=false,font=footnotesize]{subfig} \fi

\usepackage{url} 

\begin{document}
\title{On the Feasibility of Distinguishing Between Process Disturbances and Intrusions in Process Control Systems using Multivariate Statistical Process Control}

\author{\IEEEauthorblockN{Mikel Iturbe\IEEEauthorrefmark{1}, Jos\'e Camacho\IEEEauthorrefmark{2}, I\~naki Garitano\IEEEauthorrefmark{1}, Urko Zurutuza\IEEEauthorrefmark{1}, Roberto Uribeetxeberria\IEEEauthorrefmark{1}}
    \\
    \IEEEauthorblockA{\IEEEauthorrefmark{1}Department of Electronics and Computing, Faculty of Engineering \\
      Mondragon University \\
      E-20500 Arrasate-Mondrag\'on, Spain \\
      Email: \{miturbe,igaritano,uzurutuza,ruribeetxeberria\}@mondragon.edu}
      \\
  \IEEEauthorblockA{\IEEEauthorrefmark{2}
    Department of Signal Theory, Telematics and Communications -- CITIC\\
  University of Granada \\ E-18071 Granada, Spain\\
  Email: jose.camacho@ugr.es}
}

\maketitle

\begin{abstract} 
  Process Control Systems (PCSs) are the operating core of Critical Infrastructures (CIs). As such, anomaly detection has been an active research field to ensure CI normal operation. Previous approaches have leveraged network level data for anomaly detection, or have disregarded the existence of process disturbances, thus opening the possibility of mislabelling disturbances as attacks and vice versa. In this paper we present an anomaly detection and diagnostic system based on Multivariate Statistical Process Control (MSPC), that aims to distinguish between attacks and disturbances. For this end, we expand traditional MSPC to monitor process level and controller level data. We evaluate our approach using the Tennessee-Eastman process.  Results show that our approach can be used to distinguish disturbances from intrusions to a certain extent and we conclude that the proposed approach can be extended with other sources of data for improving results.
\end{abstract}


\section{Introduction} 
Process Control Systems (PCSs) are at the core of Critical Infrastructures (CIs), as they control, automate and monitor most of the processes that power modern societies. Power generation, transport, critical manufacturing, water treatment and fuel transport are some examples of CIs. As such, it is necessary to protect PCSs and related assets in order to ensure the correct functioning of modern societies.

This necessity has been further revealed by the existence of security incidents directly related to PCSs where skilled attackers disturbed normal functioning of PCSs, affecting the surrounding environment, some of them concerning CIs. 
Examples of successful cyber-attacks involving PCSs with physical impact include Stuxnet~\cite{Langner2011Stuxnet:} and the German Steel Plant incident~\cite{Bundesamt2014Die}. 

Consequently, PCS security has been the object of considerable research attention, specially in the development of novel security mechanisms. 
Among these mechanisms, Anomaly Detection Systems (ADSs) have a prominent space. 
The predictable and static nature of PCSs make them suitable candidates for anomaly detection~\cite{MitchellSurvey}. 
However, when detecting a particular anomalous event in PCSs, the factors that cause it can be diverse. 
These factors can be classified in two large sets: process disturbances or malfunctioning, and attacks or intrusions.

In this paper we analyze the limitation and possibilities of distinguishing process disturbances and intrusions by using Multivariate Statistical Process Control (MSPC) in a process agnostic manner.

The rest of the paper is organized as follows: Section~\ref{mi:sec:related} presents related works in the literature. Section~\ref{mi:sec:mspc} introduces Multivariate Statistical Process Control. Section~\ref{mi:sec:approach} outlines our approach while Section~\ref{mi:sec:results} evaluates it experimentally. Finally, Sections~\ref{mi:sec:conclusions} and~\ref{mi:sec:future} extract some conclusions and draw some lines for further work, respectively.

\section{Related Work} 
\label{mi:sec:related} 
Anomaly detection in PCSs and industrial environments in general has gathered wide attention from the scientific community.

While most of the approaches leverage network level data to detect anomalies in PCSs (see survey~\cite{MitchellSurvey}), other proposals, such as ours, address this task by leveraging process and sensor-level data.

When dealing with process level data, proposals can be further classified in two subgroups: (1) solutions that require a model of the monitored process to detect anomalies and (2) approaches where modelling the process is not necessary.
Process model dependant contributions include the work of McEvoy and Wolthusen~\cite{McEvoy2011plant} and Svendsen and Wolthusen~\cite{Svendsen2009Using}. 
While effective to detect anomalies, these approaches require accurate modelling of the physical process. 
This requirement poses an important obstacle for implementing detection systems of this nature, especially in complex processes.
More process-independent approaches on the other hand, include the work of Kiss et al.~\cite{kiss2015clustering} and Krotofil et al.~\cite{Krotofil2015Process}.

Kiss et al.~\cite{kiss2015clustering} present an anomaly detection technique based on the Gaussian mixture model clustering of the sensor-level observations.
Later, they use silhouette examinations to interpret the results.
Nevertheless, they only consider attacks as possible factors for abnormal situations in the process, without considering process faults or disturbances.
Therefore, process related anomalies could be mislabeled as attacks and vice versa.

Krotofil et al.~\cite{Krotofil2015Process} propose a method to detect when attackers tamper with sensor signals.
To this end, they use entropy to detect inconsistent sensor signals among a cluster of correlated signals.
Although they consider scenarios with process disturbances, there is no direct comparison between tampered sensor signals and similar process disturbances. 

In this approach, we go beyond the state of the art by presenting a novel security anomaly detection and diagnosis technique for PCSs. 
Additionally, we also analyze the effect of process disturbances and its effect when detecting security anomalies.

\section{Multivariate Statistical Process Control} 
\label{mi:sec:mspc}

\begin{figure}
  \centering
  \includegraphics[width=.6\columnwidth]{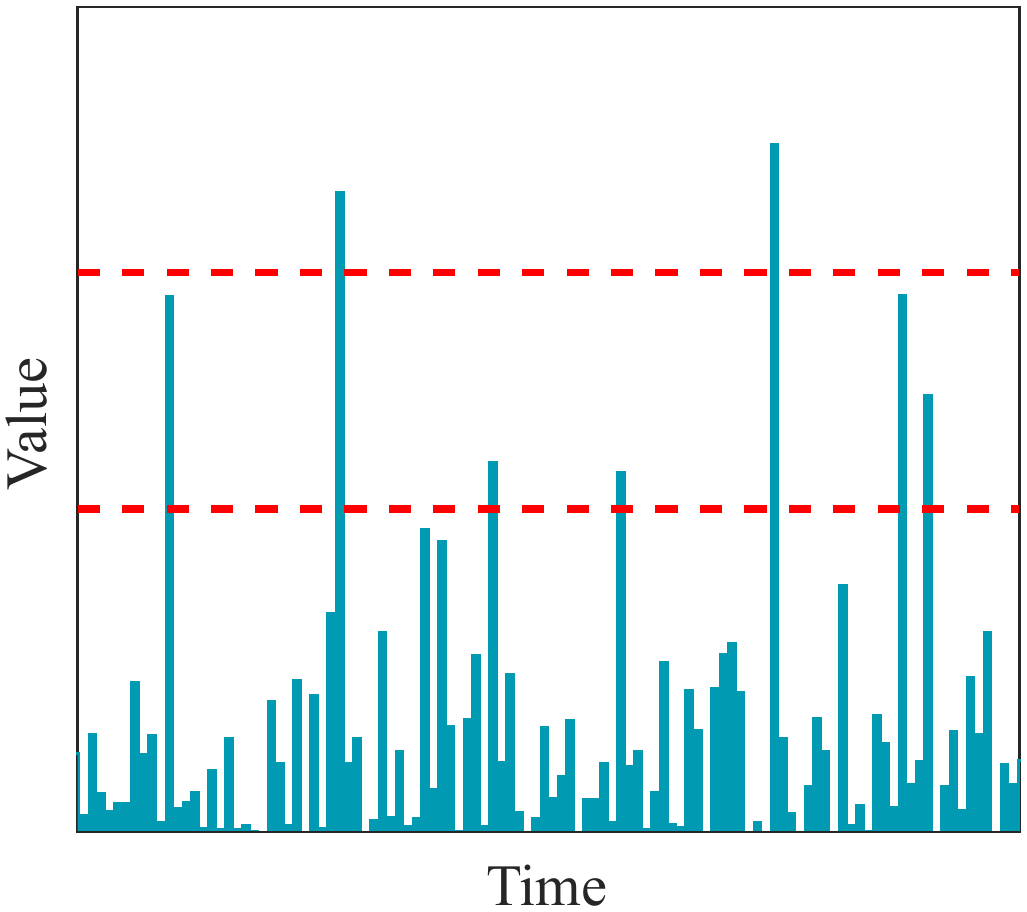}
  \caption{Example of a control chart. Control limits are presented for 95\% (lower dashed line) and for a 99\% (upper dashed line) confidence levels}
  \label{mi:fig:example_control_chart}
\end{figure}

Figure~\ref{mi:fig:example_control_chart} shows an example of a control chart. 
Under normal process operating conditions, 99\% of all the points will fall under the upper control limit.
In that case, we consider that the process is in a state of \textit{statistical control}.
It is important not to confuse the term statistical control with other similar terms, such as automatic feedback control, as they refer to different concepts.
Statistical control refers to the state of the process where only common causes of variation are present~\cite{MacGregor1995Statistical}.

The existence of consistent observation series over the established control limit, is likely to be attributed to a new special cause.
In the case of PCSs, this variation source may be attributed to attacks or process disturbances, i.e.\ an anomaly.

By using tools such as Principal Component Analysis (PCA), MSPC provides an efficient methodology to monitor variable magnitude and relation to other variables.

\subsection{PCA-based MSPC} 
Let us consider process historical data as an $\mathbf{X} = N\times M$ two-dimensional dataset, where $M$ variables are measured for $N$ observations.
PCA transforms the original $M$-dimensional variable space into a new subspace where variance is maximal.
It converts the original variables into a new set of uncorrelated variables (generally fewer in number), called Principal Components (PCs) or Latent Variables.

For a mean-centered and auto-scaled\footnote{Normalized to zero mean and unit variance} $\mathbf{X}$ and $A$ principal components, PCA follows the next expression:

\begin{equation}
  \mathbf{X} = \mathbf{T_A P^{t}_A} +\mathbf{E_A}
\end{equation}

\noindent where $\mathbf{T_A}$ is the $N \times A$ score matrix, that is, the original observations represented according to the new subspace; $\mathbf{P^{t}_A}$ is the $M \times A$ loading matrix, representing the linear combination of the original variables that form each of the PCs; finally, $\mathbf{E_A}$ is the $N \times M$ matrix of residuals. 

In PCA-based MSPC, both the scores and the residuals are monitored, each in a separate control chart~\cite{Camacho2015Multivariate}.
On the one hand, to comprise the scores, the D-statistic or Hotelling's $T^2$~\cite{hotelling1947multivariate} is monitored.
On the other hand, in the case of the residuals, the chosen statistic is the Q-statistic or $SPE$~\cite{jackson1979control}.

$D$ and $Q$ statistics are computed for each of the observations in the calibration data, and control limits are set for each of the two charts. 
Later, these statistics are also computed for incoming data and plotted in the control chart.
When an unexpected change occurs in one (or more) of the original measured $M$ variables, one (or both) of these statistics will go beyond control limits.
Thus, a $M$-dimensional monitoring scenario is effectively converted into a two-dimensional one.

In this work, we consider an event as anomalous when three consecutive observations surpass the 99\% confidence level control limit.

Once an anomaly has been detected, we use oMEDA plots~\cite{camacho2011observation} to diagnose the anomaly causes by relating anomalous events to the original variables.
In essence, oMEDA plots are bar plots where the highest or lowest values in a set of variables reflect their contribution to a group of observations. 
Therefore, when computed on a group of observations within an anomalous event, the most relevant variables related to that particular event will be the ones with the highest and lowest bars.

\section{Proposed approach}
\label{mi:sec:approach}

\begin{figure}
  \centering
  \includegraphics[width=.5\columnwidth]{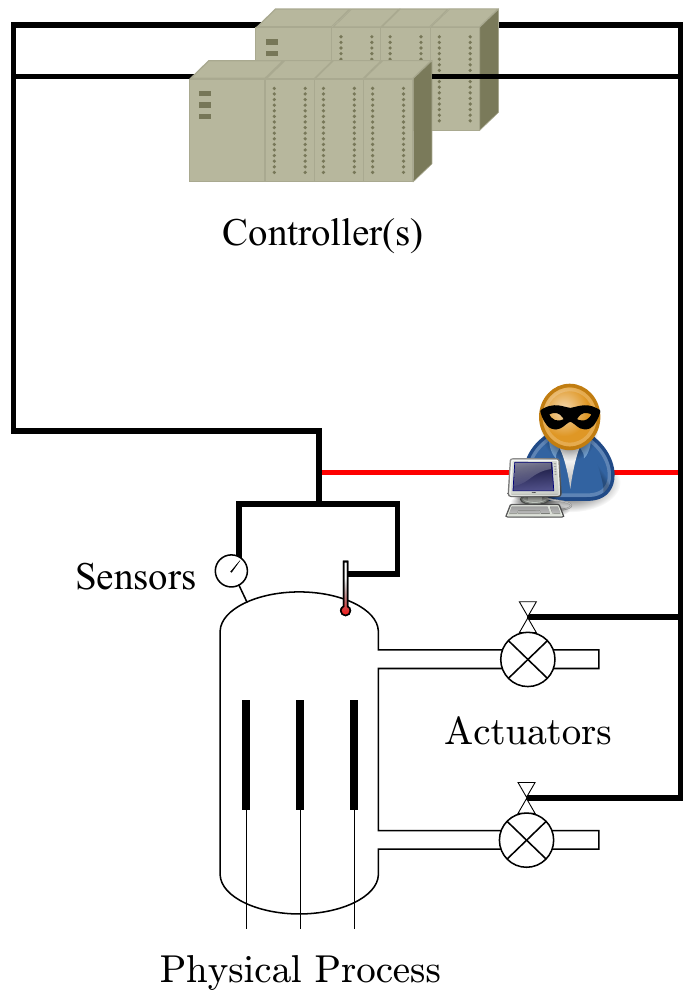}
  \caption{Example of a PCS and used attack model.}
  \label{mi:fig:example_pcs}
\end{figure}

Figure~\ref{mi:fig:example_pcs} shows an example of a PCS.
At the core of the system resides a physical process, with a fixed number of sensors and actuators. 
These sensors and actuators are the input/output devices that controllers use to interact with the process.
Controllers read process data from the sensors, and according to the control algorithm implemented in them, they decide what is the next step to be performed on the actuators.
Once the actuators change, the process evolves and with it, the sensor reading. Then, sensor  data is fed to the controllers again, thus repeating the steps.

However, the communication between process controllers and sensor/actuators is often performed over insecure transmission lines, frequently using unencrypted, unauthenticated, legacy protocols.
Thus, it is possible for an attacker to interact with the communication, performing Man-in-the-Middle (MitM) attacks. 

This can lead to situations where the data fed to the controller is not the real read by the sensors, or that the actuators receive data that was not sent as such by the controllers.

In this work we use MSPC over a simulated industrial process, the Tennessee-Eastman~\cite{downs1993plang}, to detect anomalies and diagnose their cause distinguishing between natural (disturbances) and human induced (attacks) factors.

\subsection{Tennessee-Eastman process} \label{mi:sec:te}

The Tennessee-Eastman (TE) process is a well-known challenge process, modeled after a real chemical process.

First presented by Downs and Vogel~\cite{downs1993plang}, it has been widely used by researchers to test different control strategies.
Though initially designed as a process control challenge, the TE process has also become a prominent choice among security research works~\cite{McEvoy2011plant,Cardenas2011Attacks,Krotofil2015Process,kiss2015clustering}.

In this work we use Ricker's~\cite{ricker1996decentralized} decentralized control strategy, along with the added randomness model by Krotofil et al.~\cite{Krotofil2015Process}.

The TE model has 41 measured variables (XMEAS), 12 manipulated variables (XMV) and 20 process disturbances (IDV) implemented.
For a full description of the variables and disturbances, refer to~\cite{downs1993plang}.  
The XMEAS are read by the controllers, and interact by setting values to the XMVs.
Compared to the simplified Figure~\ref{mi:fig:example_pcs}, XMEAS variables correspond to the sensor readings and XMVs to the actuator settings.
Process disturbances are unexpected and undesired changes in process conditions that can affect process normal operation. 

Out of the modelled disturbances, IDV(6) is one of the most difficult to handle. 
It models a loss of reactant in an input feed (Feed A).

The input flux of feed A is measured by XMEAS(1), whereas XMV(3) is the manipulated that controls the valve of feed A.
Therefore, it is to be expected that attacks on closing the valve XMV(3) and the existence of disturbance IDV(6), will affect similarly to XMEAS(1). 

\begin{figure}[!t] 
  \centering 
  \subfloat[IDV(6)]{\includegraphics[width=.48\columnwidth]{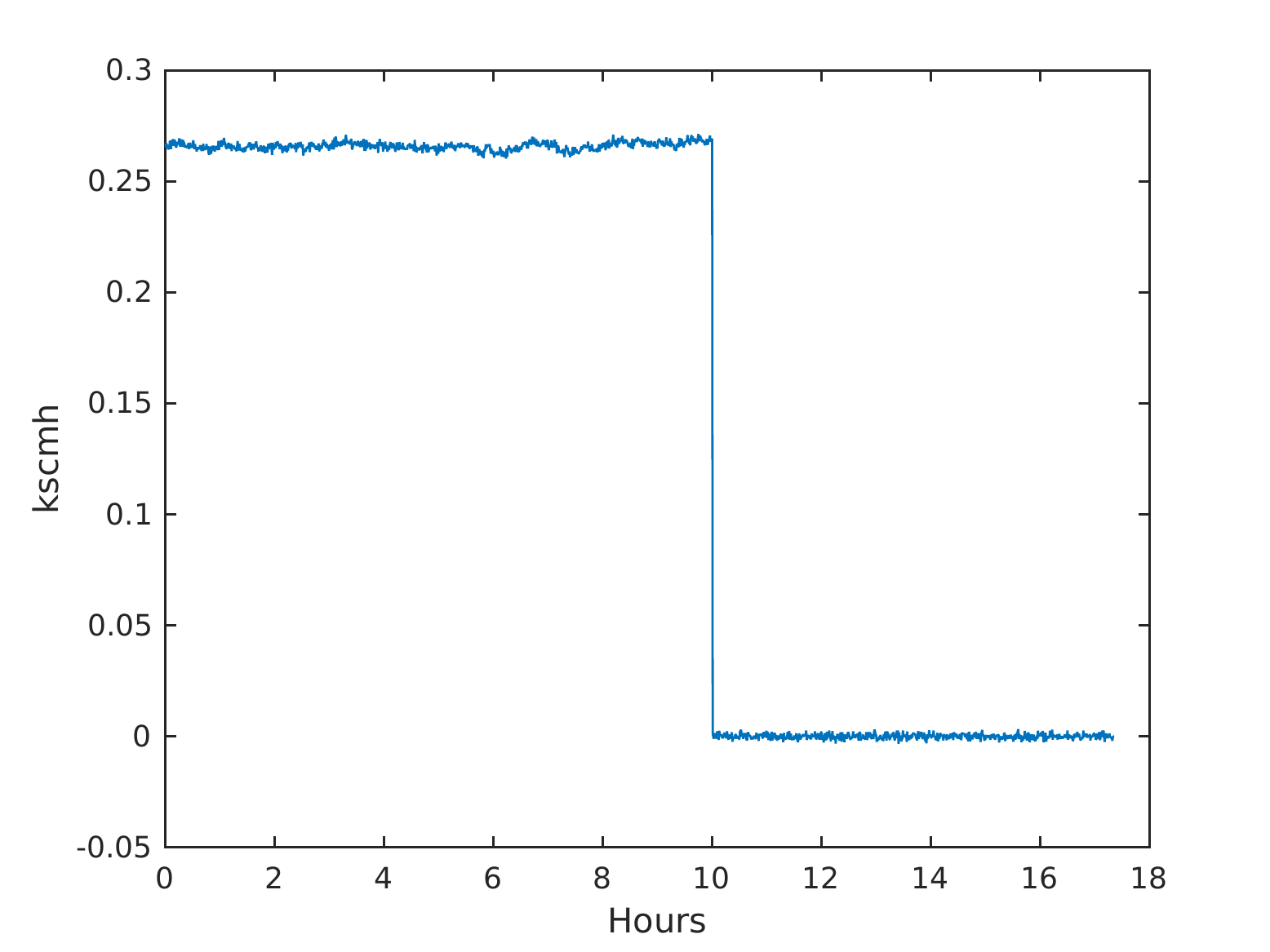}\label{mi:fig:a_feed_iv6}} \hfil
  \subfloat[Attack on XMV(3)]{\includegraphics[width=.48\columnwidth]{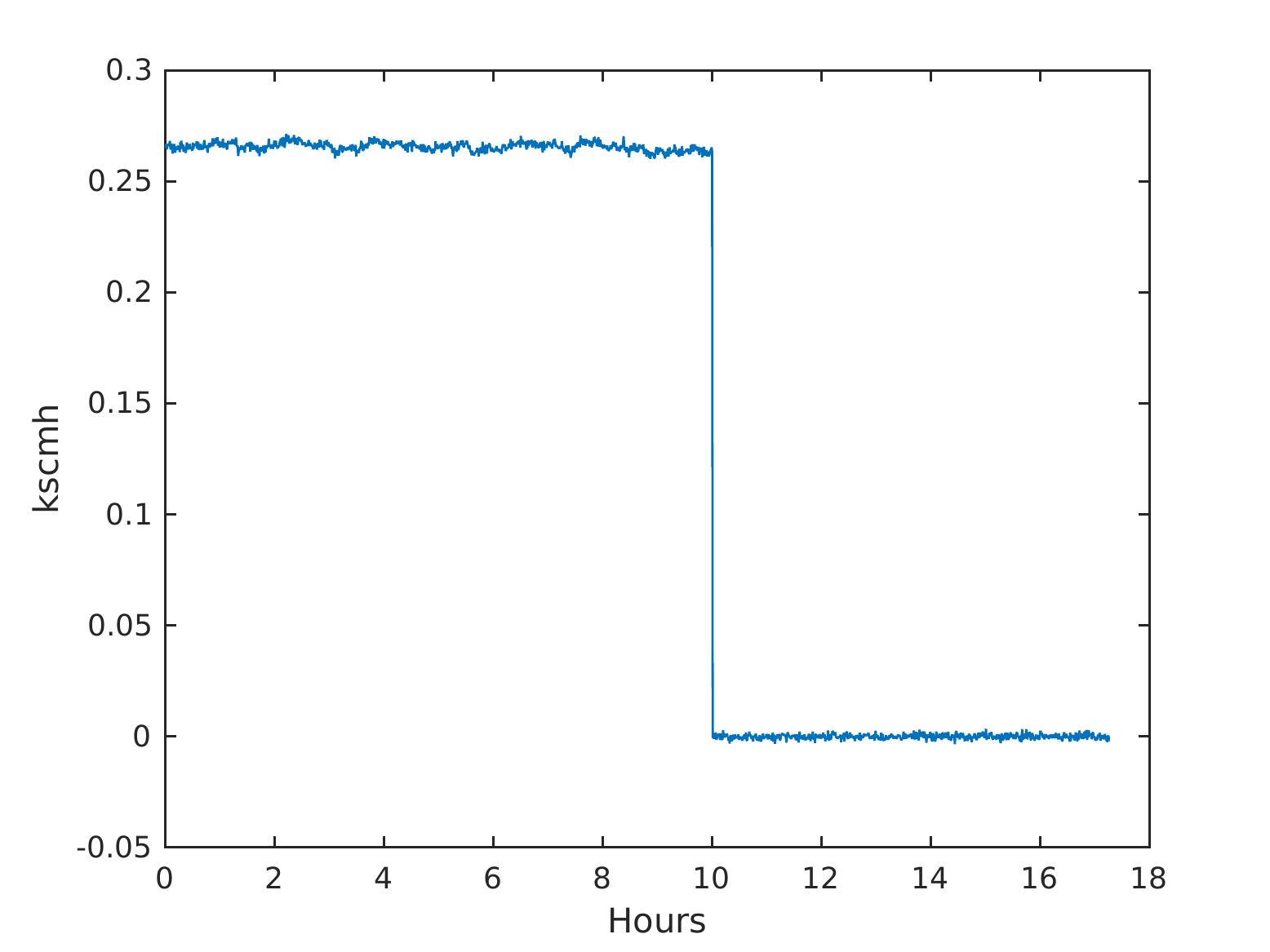}\label{mi:fig:a_feed_xmv3atk}}
  \caption{Comparison of the evolution of XMEAS(1) under disturbance IDV(6) or an integrity attack on XMV(3).} \label{mi:fig:a_feed} 
\end{figure}

Figure~\ref{mi:fig:a_feed} shows both situations.
When monitoring XMEAS(1), there is almost no difference between IDV(6) and an integrity attack on XMV(3) where the attacker commands closing the valve controlling feed A, as the flow decreases abruptly in both cases.
Both the disturbance and the attack occur at the tenth hour.
After 17 hours and 43 minutes, the process shuts down in both cases as the stripper liquid level becomes too low to continue safe operation of the plant. 

Having a process disturbance and a potential attack on a process variable that react almost identically with the process provides a sound setup to test the performance of techniques that try to distinguish them.

\subsection{Adversary modelling}
The adversary and attack models considered in this scenario are the ones proposed by Krotofil et al.~\cite{Krotofil2015Process}.

We consider that the adversary is able to read and manipulate network traffic, between controllers and the physical process as depicted in Figure~\ref{mi:fig:example_pcs}.

Therefore, the attacker is capable of manipulating input data both at the controllers' (forged XMEAS data) and/or the physical process' (forged XMV data) end, performing an integrity attack.

Following the model of Krotofil et al.~\cite{Krotofil2015Process}, we consider an attacked variable $Y'_i(t)$  at time $t$, $0\leq t\leq T$ as follows, where $T$ is the duration of the simulation and $T_a$ the arbitrary attack interval. An integrity attack is defined as follows:

\begin{equation}
  Y'_i(t) = 
  \begin{cases}
  Y_i(t), & \mbox{for } t \notin T_a \\ 
  Y^{a}_i(t), & \mbox{for } t \in T_a 
  \end{cases} 
\end{equation}

\noindent where $Y_i^a(t)$ is the modified variable value injected by the attacker.

Similarly, during DoS, the attacker effectively stops communication, and no communication reaches the actuator or the controller. Krotofil et al.~\cite{Krotofil2015Process} define as a DoS attack starting at $t_a$ as:

\begin{equation}
  Y^a_i(t)= Y_i(t_a-1)
\end{equation}

\noindent where $Y_i^a$ is the last value received before the DoS attacks.

\section{Experimental results}
\label{mi:sec:results} 

In order to evaluate our approach, we conduct a set of experiments where the
randomized TE model is run ten times per anomalous situation. 
The model we used for the set of experiments is the DVCP-TE
model presented by Krotofil and Larsen~\cite{krotofil2015}, freely
available on Github\footnote{\url{http://github.com/satejnik/DVCP-TE}}.
The time length of each simulation is 72 hours, except in the cases where the process shut itself down due to safety constraints. 
For each simulation hour, variable data is recorded 2000 times, that
is, every 1.75 seconds.
Calibration data consists of 30 runs, and this data is used to build the MSPC
model and establish the control limits of the $D$ and $Q$ statistics.

All anomalies start at the 10th hour of simulation.
For each of the anomalous situations, we calculate the Average Run Length (ARL), that refers to the lapsed time between the start of the anomalous event and its detection in the control charts.
As previously stated, an event is flagged as anomalous when three consecutive observations surpass the 99\% control limit.

Once an anomaly is flagged, oMEDA charts are computed for the set of the first observations that surpass control limits in each of the ten runs in either of the two control charts (monitoring $D$ and $Q$-statistic).

For each anomalous event two plots are created, one with real process data (data the process receives and sends), and the other with controller level data. 
Both data sets will be identical in case of an attack free environment.
But, in the case of attacks, both data sets will diverge.

For the analysis of the process data, and plotting purposes, we used the MEDA toolbox~\cite{Camacho2015Multivariate}.

We set four different scenarios: \textit{a)} Disturbance IDV(6), \textit{b)} Integrity attack on XMV(3), \textit{c)} Integrity attack on XMEAS(1), and \textit{d)} Denial of Service on XMV(3).

\begin{figure*}[!ht] 
  \centering 
  \subfloat[IDV(6)]{\includegraphics[width=.7\columnwidth]{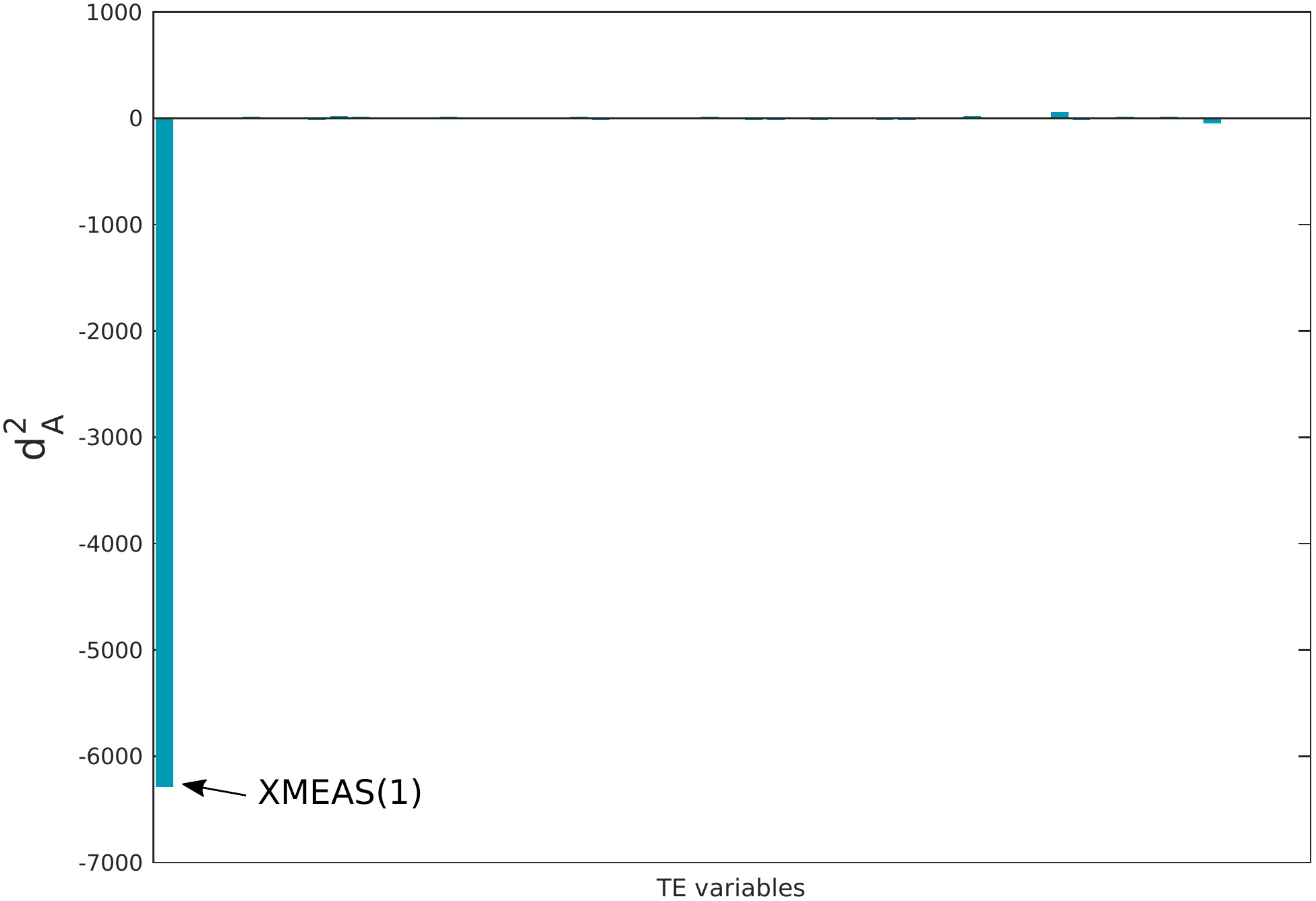}\label{mi:fig:omeda_idv6_controller}}
  \subfloat[Integrity attack on XMV(3)]{\includegraphics[width=.69\columnwidth]{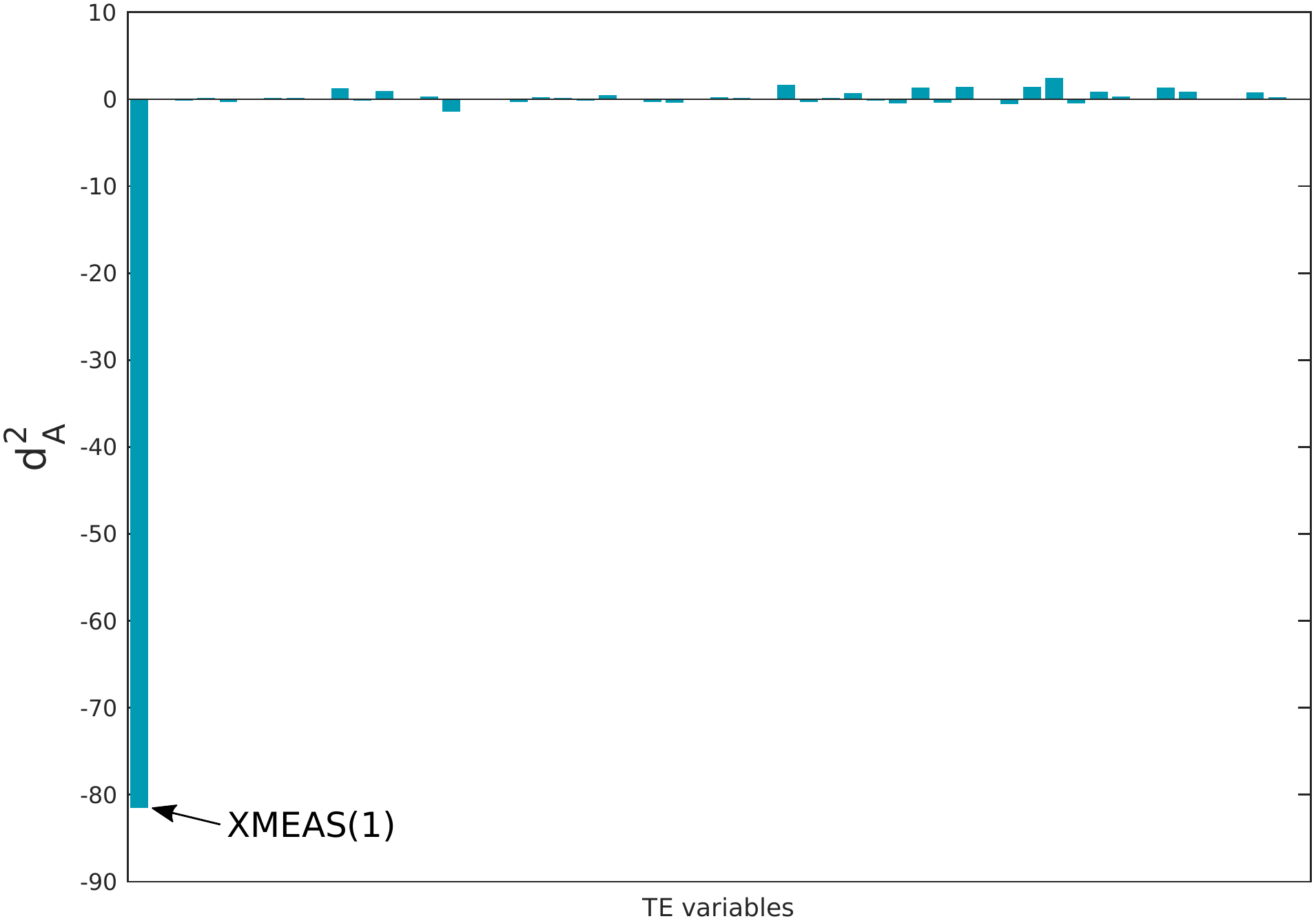}\label{mi:fig:omeda_xmv3_close_controller}}\\
  \subfloat[Integrity attack on XMEAS(1)]{\includegraphics[width=.7\columnwidth]{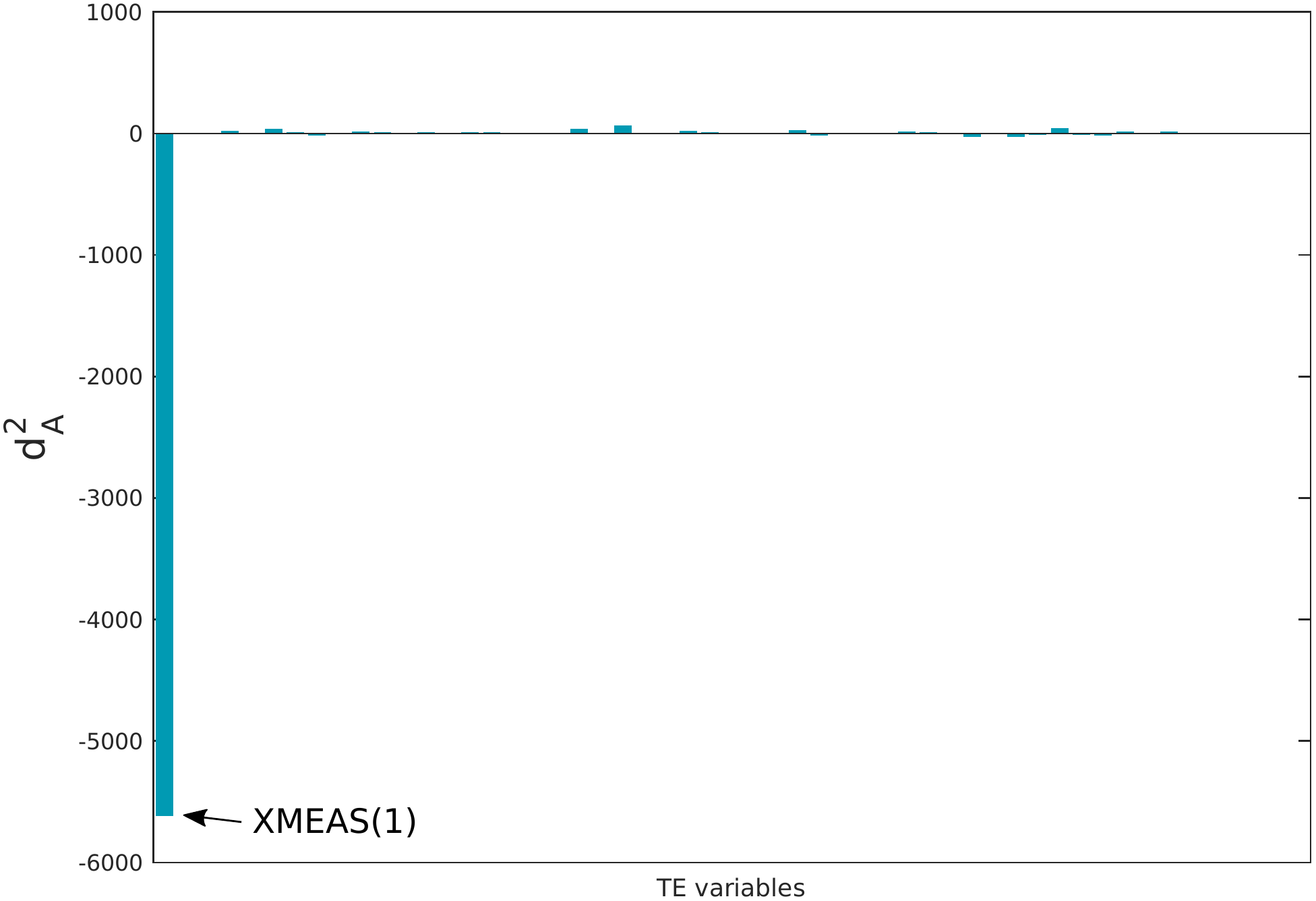}\label{mi:fig:omeda_xmeas1_close_controller}}
  \subfloat[DoS attack on XMV(3)]{\includegraphics[width=.69\columnwidth]{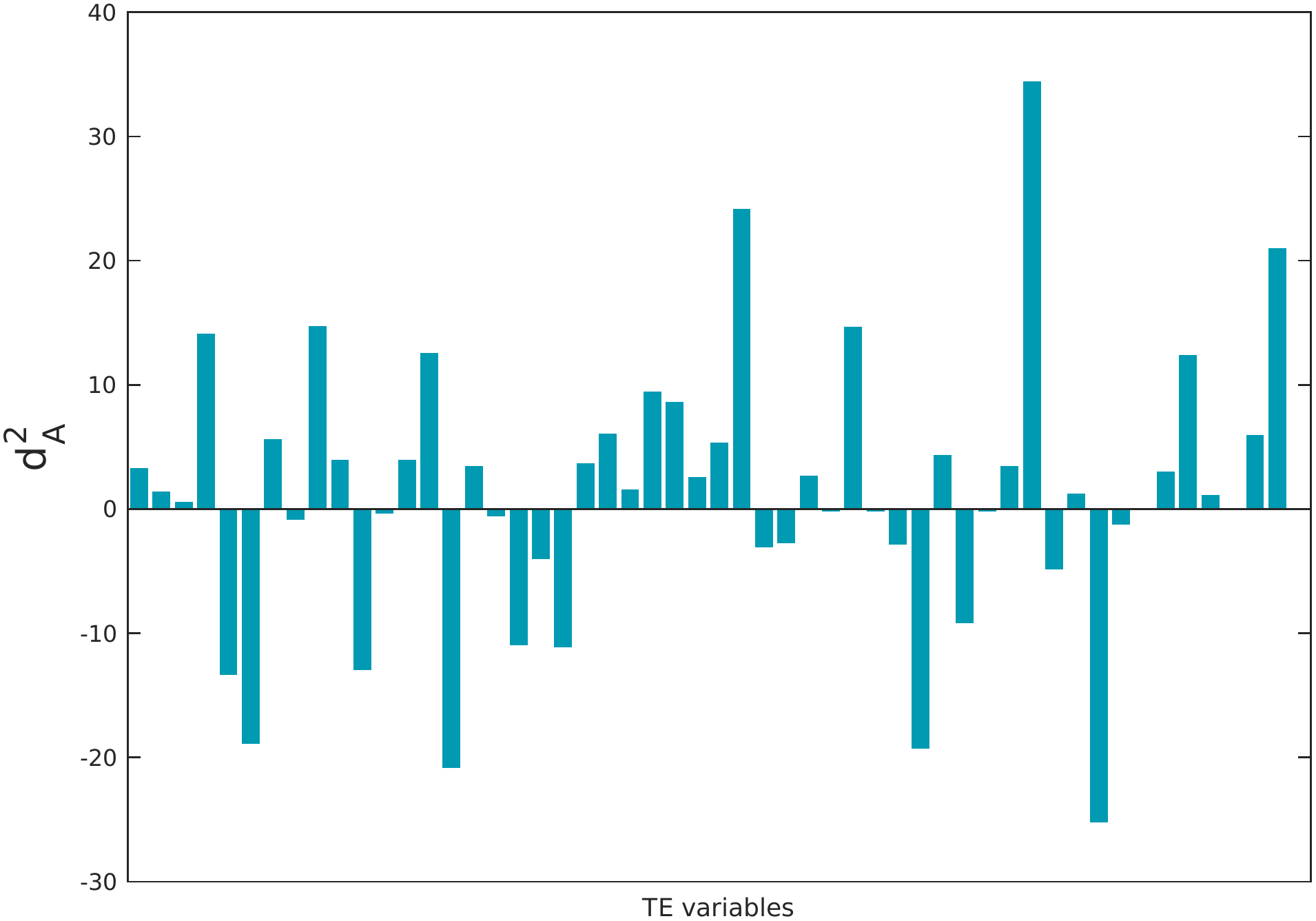}\label{mi:fig:omeda_xmv3_dos_controller}}
  \caption{oMEDA plots of different anomalies from the controller point of view} \label{mi:fig:cn_omedas} 
\end{figure*}

\begin{figure*}[!ht] 
  \centering 
  \subfloat[IDV(6)]{\includegraphics[width=.7\columnwidth]{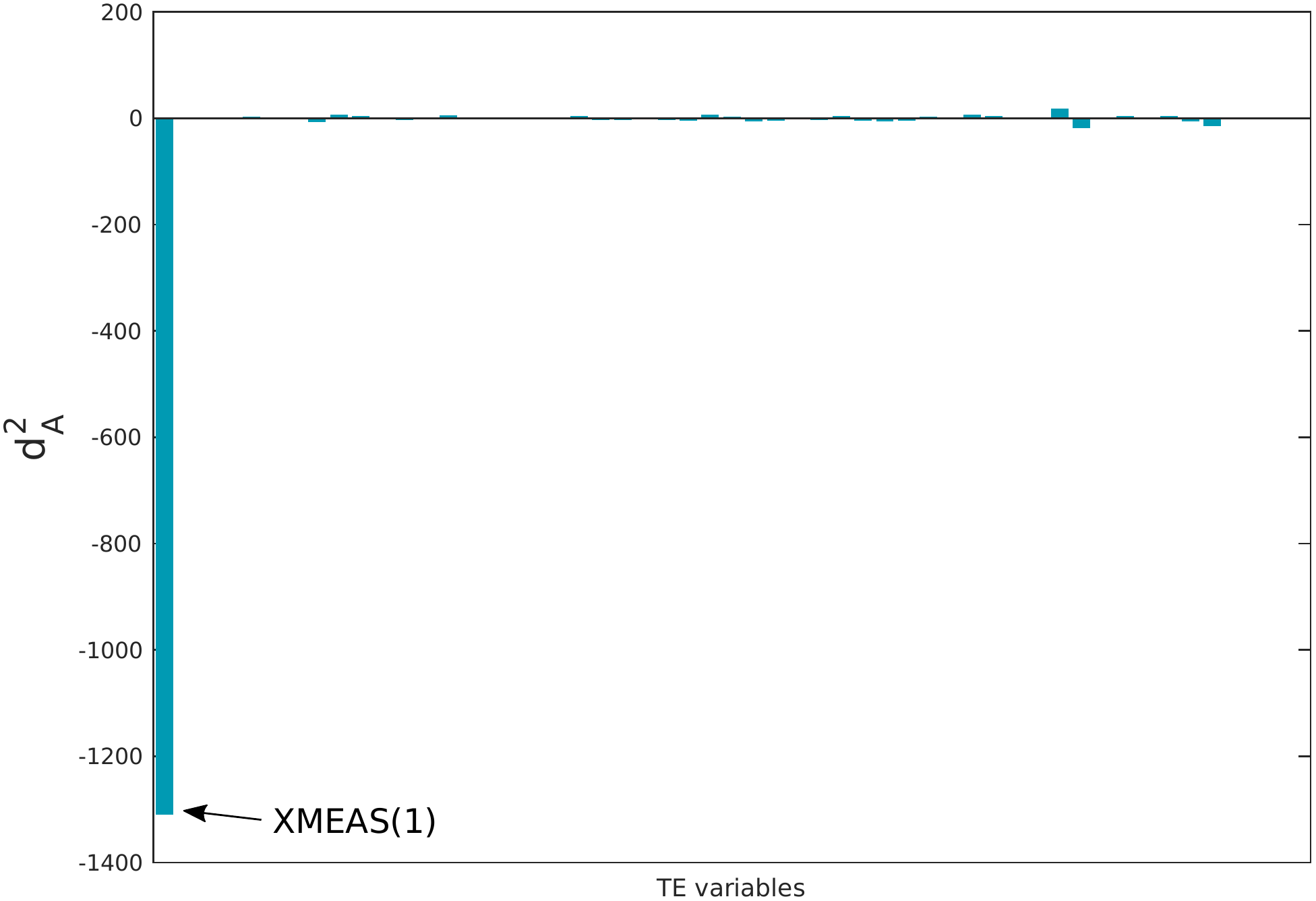}\label{mi:fig:omeda_idv6_process}}
  \subfloat[Integrity attack on XMV(3)]{\includegraphics[width=.7\columnwidth]{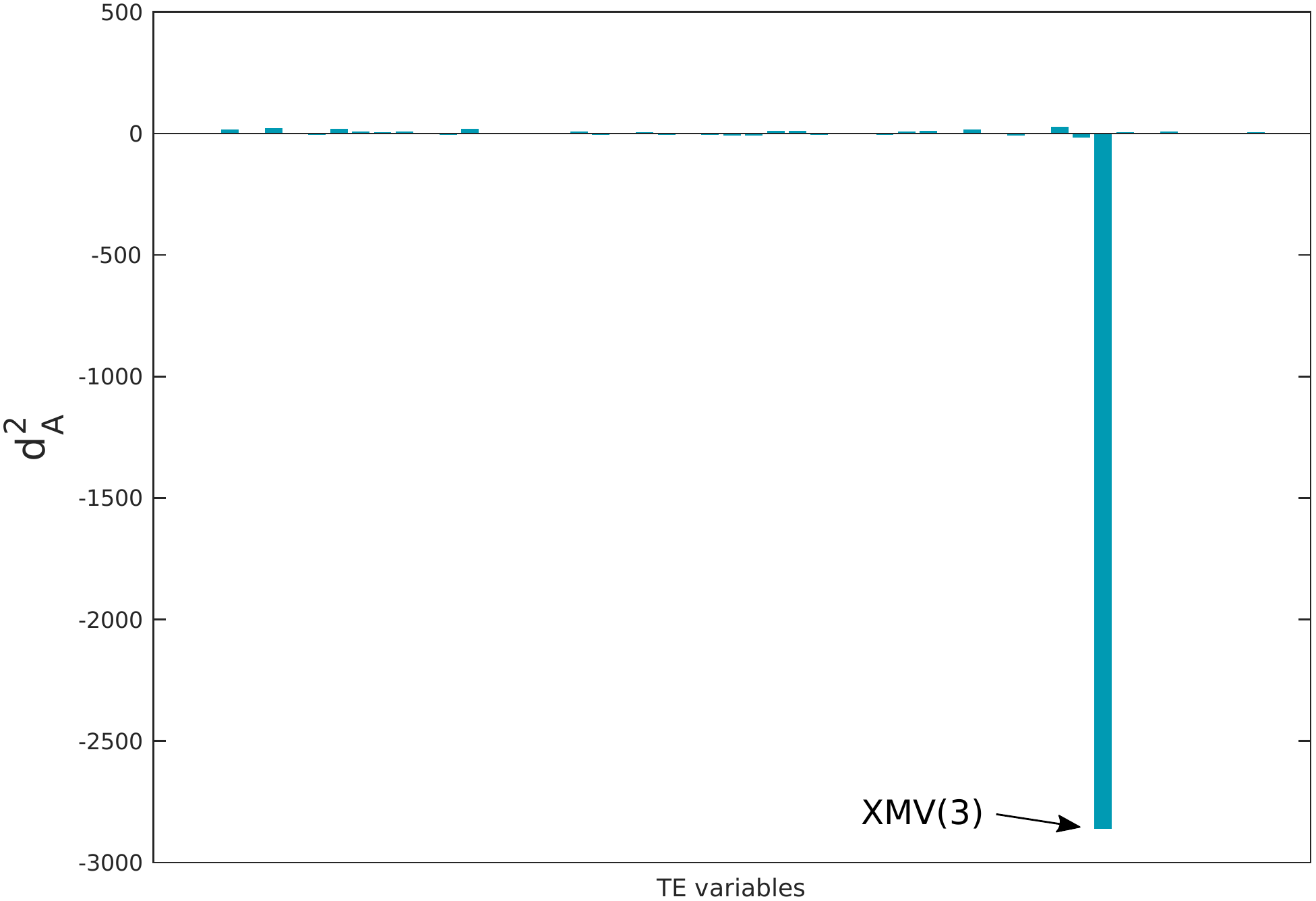}\label{mi:fig:omeda_xmv3_close_process}}\\
  \subfloat[Integrity attack on XMEAS(1)]{\includegraphics[width=.7\columnwidth]{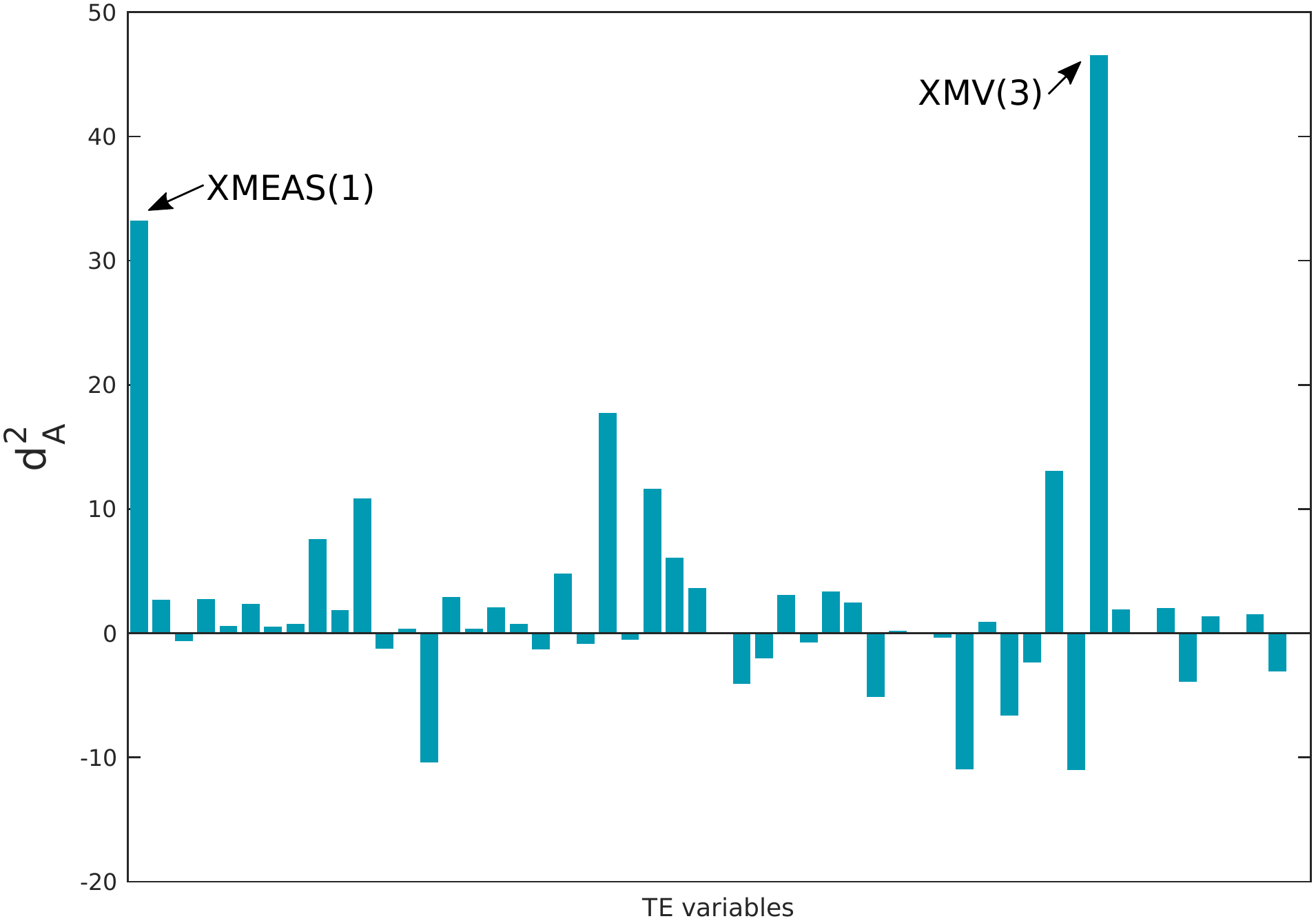}\label{mi:fig:omeda_xmeas1_close_process}}
  \subfloat[DoS attack on XMV(3)]{\includegraphics[width=.7\columnwidth]{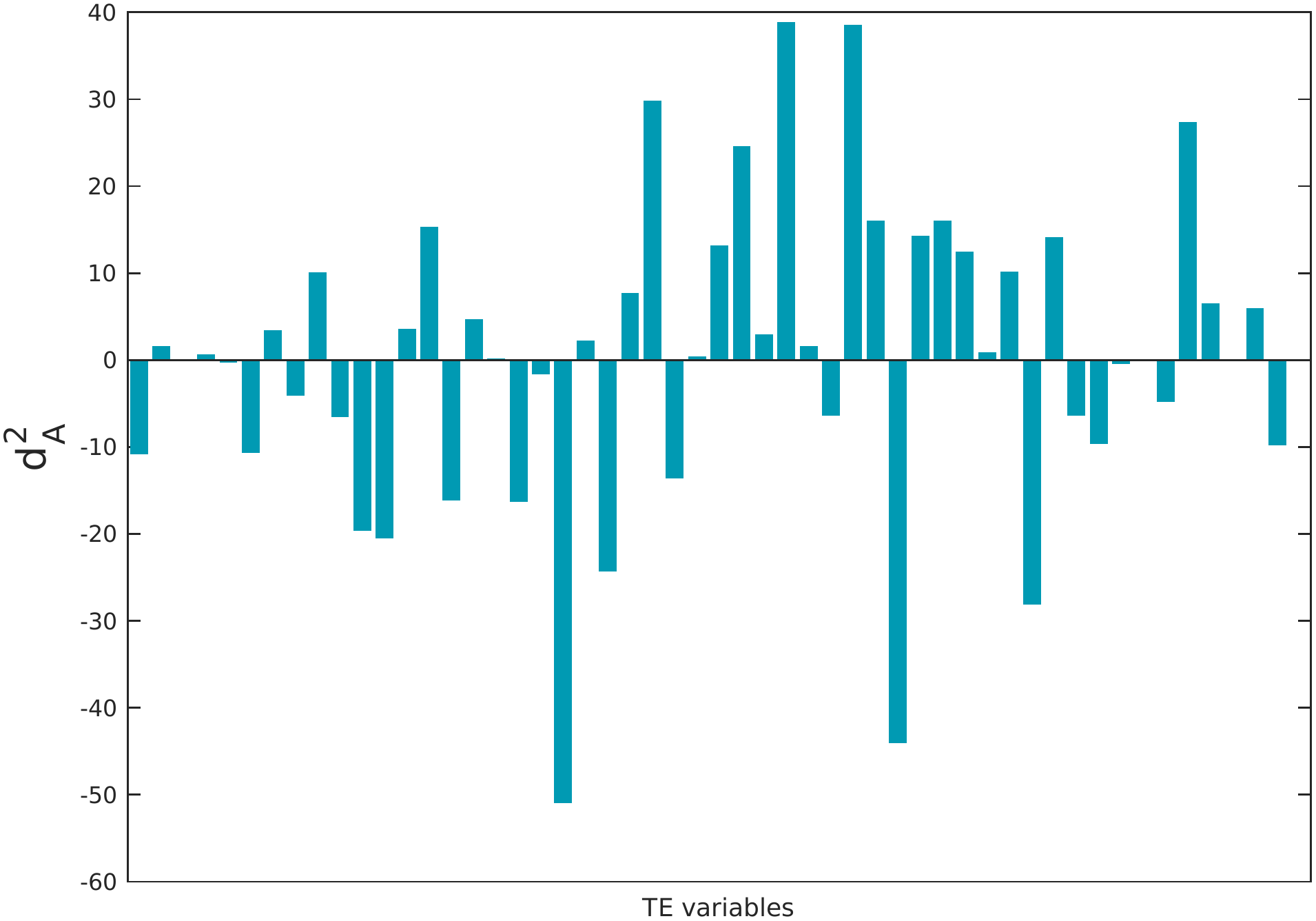}\label{mi:fig:omeda_xmv3_dos_process}}
  \caption{oMEDA plots of different anomalies from the process point of view} \label{mi:fig:sn_omedas} 
\end{figure*}

Resultant oMEDA for controller level and process level variables are shown in Figures~\ref{mi:fig:cn_omedas} and~\ref{mi:fig:sn_omedas}, respectively.

Figures~\ref{mi:fig:omeda_xmv3_close_controller} and~\ref{mi:fig:omeda_xmv3_close_process} show the oMEDA charts for the case of where the attacker performs and attack and closes the valve of feed A.
In this case, from the controllers point of view, the anomaly is similar to the one with IDV(6). 
It is when we look at process-level data that we see that the real concerned variable is not XMEAS(1).
Rather, the attacker is manipulating XMV(3) to perform the attack.

Figures~\ref{mi:fig:omeda_xmeas1_close_controller} and~\ref{mi:fig:omeda_xmeas1_close_process} shows the oMEDA plots of an scenario where the attacker manipulates the XMEAS(1) variable and sets it to zero.
Therefore, the controller receives the information that there is no flow in Feed A.
That is why the XMEAS(1) value from the controller point of view is lower than usual, because the attacker has set it so.
As the control algorithm tries to tackle the situation, it opens XMV(3) more, and thus flowing more reactant A to the process.
From the process point of view, that is the reason XMV(3) and XMEAS(1) have higher values than usual.

oMEDA plots for a DoS attack on XMV(3) are shown in Figures~\ref{mi:fig:omeda_xmv3_dos_controller} and~\ref{mi:fig:omeda_xmv3_dos_process}.
In this scenario, the process keeps receiving a constant value, previous to the attack.
Neither of the oMEDA plots show a variable, let alone XMV(3) that stands out clearly among others.
It is worth noting that the ARL, on the other cases almost immediate, in DoS attacks is significantly higher. 
In this case, DoS detection takes almost an hour.

\subsection{Discussion}
Our approach detects all anomalous situations of disturbances and attacks. 

However, when diagnosing an anomaly, controller-based readings --on witch traditional MSPC has relied on-- are not enough to do so correctly.
Both integrity attacks and the process disturbance are diagnosed in a very similar way, in a manner that it is not feasible to distinguish what caused the anomaly.

To address this matter, we have extended the MSPC model and measure both process and controller level variables.
Having this two level input makes possible monitoring a wide range of process variables.
Moreover, when anomalies occur, it is possible to distinguish the origin of the anomaly.

When considering DoS attacks, the ARL is significantly longer than with integrity attacks or process disturbances and the diagnosis is not as clear as in the other scenarios. 

\section{Conclusions} 
\label{mi:sec:conclusions} 
We have presented a process-independent approach to detect and distinguish process
disturbances from related attacks. Unlike previous approaches, it is not a
process-dependant approach and it is able distinguish between disturbances and
attacks.

Our methodology is based on MSPC for anomaly detection and oMEDA plots for anomaly diagnosis.
We have used the popular Tennessee-Eastman process to experimentally evaluate our approach.

Distinguishing process disturbances and low level attacks in PCSs is a complex task, especially if all controller's I/O are to be considered compromised.

We extended the traditional MSPC model to monitor both controller and process level variables, to efficiently monitor PCSs.
Often, PCSs assign measured variables to the manipulated ones, so this approach is feasible in these environments.
This scenario, would also complicate the work of an attacker, as it would need to forge both the target manipulated variable and the associated measured one to avoid detection.

When analyzing process disturbances or integrity attacks, the oMEDA plots clearly show the implicated variables. 
In the case of DoS, detection time is significantly longer and the diagnosis with oMEDA is not related to the attacked variable.

\section{Future work} 
\label{mi:sec:future} 
To overcome current anomaly diagnosis limitations, it is necessary to add more information to the MSPC model. 

In the case of PCSs, a promising source of additional information is the one
created at the network level (packets, flows, logs etc.).

MSPC-like methodologies have already been used in regular IT networked environments for security monitoring~\cite{Camacho2016pca}.

We are confident that adding network-level variables to the ones of the process will ease anomaly diagnosis (e.g.\ by detecting increased traffic in the case of network DoS attacks) and will also shorten the ARL required to detect anomalies, as while the process might be slow to surpass control limits due to slow dynamics, network variables show more immediate information.

\section*{Acknowledgments}
Research in this paper is partially supported by the Basque Government's Elkartek program under grant number KK-2015/0000080, the Provincial Council of Gipuzkoa's ``Red guipuzcoana de Ciencia, Tecnolog\'ia e Innovaci\'on'' program through grant 56/15 and the Spanish Government-MINECO (Ministerio de Econom\'ia y Competitividad) and FEDER funds, through project TIN2014-60346-R.

\bibliographystyle{IEEEtran}
\bibliography{references_miturbe}
\end{document}